\documentclass[11pt,lscape]{article}
\usepackage{graphics}
\usepackage{graphicx}
\usepackage{lscape}
\usepackage{epsfig}
\usepackage{amsmath}
\usepackage{amssymb}
\usepackage{subcaption}

\author{
  Deniz Yenig{\"u}n\\
  Istanbul Bilgi University \\ Istanbul, Turkey\\
  \texttt{deniz.yenigun@bilgi.edu.tr}
  \and
  G{\"u}ne{\c s} Ertan\\
  Ko\c{c} University \\ Istanbul, Turkey\\
  \texttt{gunesertan@ku.edu.tr}
   \and
  Michael Siciliano\\
  University of Illinois \\ Chicago, IL, USA \\
  \texttt{sicilian@uic.edu}
}
\title{Omission and Commission Errors in Network Cognition and Network Estimation using ROC Curve}

\begin{document}

\maketitle

{ \small \begin{abstract}
\noindent Cognitive Social Structure (CSS) network studies collect relational data on respondents' direct ties and their perception of ties among all other individuals in the network. When reporting their perception networks, respondents commit two types of errors, namely, omission (false negatives) and commission (false positives) errors. We first assess the relationship between these two error types, and their contributions on the overall respondent accuracy. Next we propose a method for estimating networks based on perceptions of a random sample of respondents from a bounded social network, which utilizes the Receiving Operator Characteristic (ROC) curve for balancing the tradeoffs between omission and commission errors. A comparative numerical study shows that the proposed estimation method performs well. This new method can be easily integrated to organization studies that use randomized surveys to study multiple organizations. The burgeoning field of multilevel analysis of inter-organizational networks can also immensely benefit from this approach.

\vspace{0.8cm}
\end{abstract}
}

\pagenumbering{arabic}
\setcounter{page}{1}
\section{Introduction}\label{intro}

Most individual level network oriented studies ask respondents to recall their own direct social ties. Cognitive Social Structure (CSS) based studies on the other hand, collect data not only on the actor's own relations, but also his/her perception of social ties among all other actors in the network. In complete CSS designs for networks of size $N$, each respondent is required to answer questions about $N^{2}-N$ possible relations in the network. Formalized by David Krackhardt (1987), CSS scholarship mostly centers around determinants of accuracy of network cognition such as personality traits, (Casciaro, 1998; Casciaro et al. 1999), power (Krackhardt, 1990; Simpson et al. 2011), need for closure and homophily (Flyn et al, 2010), social exclusion (O'Connor and Gladstone, 2015), network centrality (Krackhardt, 1987; Grippa and Gloor, 2009; Simpson and Borch, 2005), and egocentric biases (Kumbasar et al., 1994; Johnson and Orbach, 2002).

This paper deviates from traditional CSS studies and aims at (1) presenting a tool for CSS researchers to classify and analyze different errors of perception, and (2) providing a new network estimation method that is designed to minimize overall errors while controlling for different error types. The method is based on randomly selected CSS slices (actors' perception matrices) and therefore does not require high response rates to estimate the network structure. The proposed density weighted Receiving Operator Characteristic (ROC) curve based method is shown to produce network estimates with relatively low errors. Moreover, this new method provides a decision tool for researchers for determining the threshold $k$ (the minimum required number of actors claiming the presence of a tie to assign a tie between two actors in the estimated network) that enables them to visually inspect the distribution of errors at every level of $k$. The recognition of error variation in data collection is likely to benefit network scholars in areas in which costs of making an error of omission or commission may be significantly different.

The rest of this paper is organized as follows: In Section \ref{types} we introduce the five data sets we consider and present summaries of different error types in individuals' perceptions. In Section \ref{estimation} we propose a network estimation method controlling these error types. Section \ref{numerical} reports the results of a numerical study for evaluating the performance of the proposed estimation methodology, and Section \ref{conclusion} concludes. The paper is accompanied by an R package, \texttt{cssTools} (Yenig\"un et al. 2016), that allows users to manage and dissect CSS data and to estimate networks based on random samples of CSS slices.

\section{Error Types in CSS}\label{types}

To date most CSS based studies conceptualized errors as absent or present, only very few studies looked at the specific type of perception errors, and to our knowledge none compared the relationship between error types. For example, based on an experimental study O'Connor and Gladstone (2015) found that experience of social exclusion leads to false positives, in other words, errors of commission. Flyn et al. (2010) showed that need for closure is also associated with false positives due to respondents' tendency to perceive transitive ties. In another experimental study Dessi et al. (2016) showed that individuals tended to underestimate mean degree of the networks that they were asked to recall during the experiments.

Unlike this line of research, we first conceptualize and assess cognitive errors as Type 1 (errors of commission, or false positives) and Type 2 (errors of omission, or false negatives), and examine the relationship between the two types of cognitive inaccuracy and their association with overall respondent reliability. To be more specific, we define Type 1 error as an instance of error when an individual perceives there to be a tie between two actors when in fact there is no tie, and Type 2 error as an instance when perception says there is no tie when in fact there is a tie.

We use the following five data sets to analyze the errors and test our estimation technique: (1) High Tech Managers: 21 managers in a machinery firm (Krackhardt, 1987); (2) Silicon Systems: 36 semiskilled production and service workers from a small entrepreneurial firm (Krackhardt, 1990); (3) Pacific Distributers: 33 key personnel from the headquarters of a logistics company (Krackhardt and Kilduff, 1990); (4) Government Office: 36 government employees at the federal level (Krackhardt and Kilduff, 1999); and (5) Italian University: 25 researchers across partner research centers at a university (Casciaro, 1998).

All five data sets are collected according to the CSS method, where each actor not only reports his or her self-ties, but also answers questions on all possible dyads in the network. A CSS for a network involving $N$ individuals is usually represented by a three dimensional array $R_{i,j,m}$ ($i,j,m=1,...,N$), where $i$ is the sender, $j$ is the receiver, and $m$ is the perceiver of the relationship (Krackhardt, 1987). As mentioned avbove, an actor's network perception matrix is referred to as a CSS slice. Based on the work of Krackhardt (1990), for a given CSS array we construct the true network using the locally aggregate structures (LAS) intersection rule, according to which a tie is considered to be present in the true network if and only if both parties in a dyad report the tie. Note that these ties considered in the networks are directional, and thus for the $i$-$j$ tie to exist, $i$ must claim to send a tie to $j$, and $j$ must claim to receive a tie from $i$. After constructing the true network in this fashion, we explore the different types of errors in perceptions.

\begin{table}[h]
\begin{center}
\begin{scriptsize}
\centering
\begin{tabular}{lcccccc}
&&\multicolumn{2}{c}{Type I Error Rate}&\multicolumn{2}{c}{Type II Error Rate} & \\ 
Data Set & $N$ & Mean & St. Dev. & Mean & St. Dev. & Type I \& II Correlation \\ \hline
Italian University    &    24    &    0.087    &    0.090    &    0.539    &    0.179    &    -0.68    \\
High Tech Managers    &    21    &    0.052    &    0.049    &    0.636    &    0.174    &    -0.77    \\
Silicon Systems    &    36    &    0.136    &    0.140    &    0.641    &    0.204    &    -0.82    \\
Government Office    &    36    &    0.052    &    0.073    &    0.723    &    0.243    &    -0.94    \\
Pacific Distributers    &    48    &    0.028    &    0.029    &    0.701    &    0.176    &    -0.86    \\ \hline
\end{tabular}
\end{scriptsize}
\caption{Summary statistics for error rates across five data sets.}\label{tab:Summary}
\end{center}
\end{table}

For all five data sets, Table \ref{tab:Summary} displays the mean and standard deviation of both error rates, as well as the correlation between them. Error rates are calculated based on the ratio of the frequency of the particular error type for each actor to the possible number of each error type. Type 1 error rates are considerably low in comparison to Type 2 rates. Average Type 1 error rates range between 0.028 and 0.136 (with standard deviations ranging between 0.029 and 0.140) while average Type 2 error rates range between 0.539 and 0.723 (with standard deviations ranging between 0.174 and 0.243). This difference can be explained by the low true network densities, all of which are less than 0.2. In a sparse network even a small number of omission errors lead to a large rate of Type 2 error since overall number of ``1's'' in the true adjacency matrix is very small. However, committing errors of commission in similar amounts will translate into lower Type 1 error rates since there is a large number of ``0's'' in the adjacency matrix. Since majority of social networks tend to have low density (Anderson et al. 1999), these patterns are likely to be observed in many other contexts as well.

\begin{figure}[h]
\begin{center}
\includegraphics[scale=0.7]{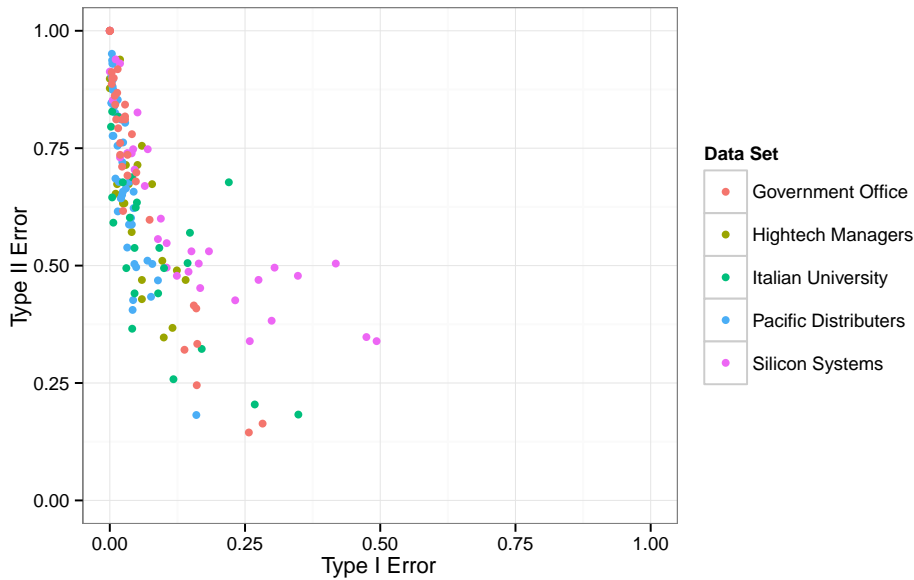}
\caption{Type 1 and Type 2 error rates across five data sets.}
\label{fig:errorRate}
\end{center}
\end{figure}

The two error types have strong negative correlations in all five data sets, which is also visible from Figure \ref{fig:errorRate}. In this scatter plot, each actor in each data set is represented by a single point, and a good mix of colors indicating data sets reveal that the five sets investigated share common characteristics in terms of error types. The strong negative correlations suggest that for most individuals, their overall errors are dominated by either Type 1 or Type 2 error.

\begin{figure}[h]
\begin{center}
\includegraphics[scale=0.5]{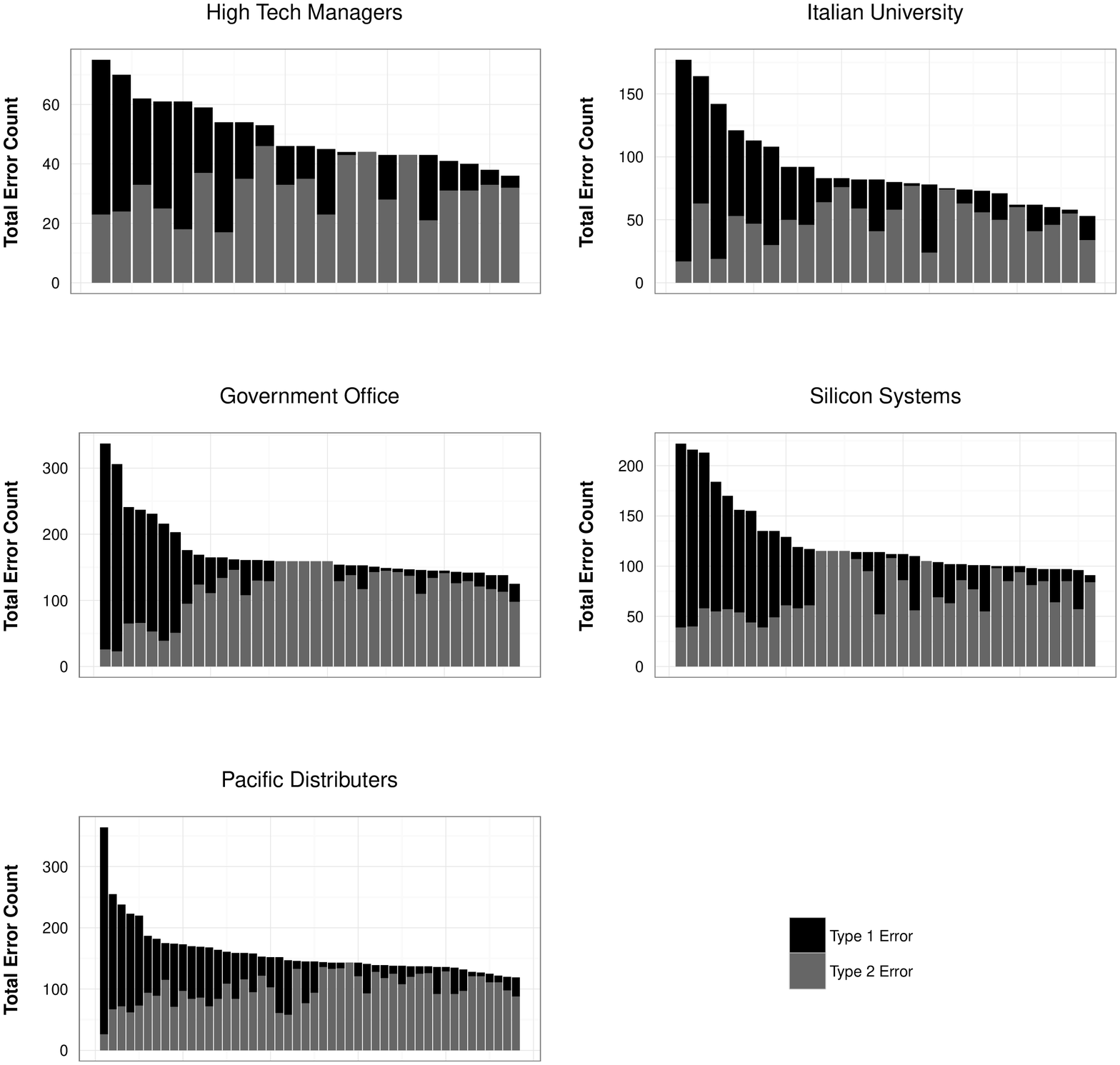}
\caption{Individuals' Type 1 and Type 2 error counts across five data sets. Each bar represents an individual.}
\label{fig:errorCount}
\end{center}
\end{figure}

Figure \ref{fig:errorCount} exhibits the breakdown of Type 1 and Type 2 error counts for overall perception error of each individual actor. Each vertical bar represents the total error count of an individual in the network, and colors indicate error types. For each data set, when we classify individuals as {\em low error} and {\em high error} in terms of total error count, we see a similar pattern. In the low error group Type 2 errors dominate the overall error, with only a few Type 1 errors. In the high error group both error types are observed, and the ratio of Type 1 errors to Type 2 errors significantly increase as the total error count increases. These plots suggest that the low error group consists of more conservative individuals who do not report a tie unless they are confident (or aware), which creates many Type 2 errors. However, since they do not have the tendency to report a non existing tie, they commit very few Type 1 errors. The high error group consists of more ``liberal'' individuals who have the tendency to report many ties without being picky, producing many Type 1 errors. On the other hand, since they do not have the tendency to reject ties, their Type 2 errors decrease. This interpretation of high and low error groups also explains the negative correlations between error types. We also note that the prominence of Type 1 errors in the high error group can also be explained by the low density prevalence. Since there is only limited number of ``1's'' in the true network, individuals with extreme levels of errors need to perceive many ``0's'' as ``1's'', hence commit large numbers of Type 1 error.

In sum, scholarly studies on network cognition so far mostly focused on determinants of cognitive accuracy and ignored different types of cognitive errors. A closer inspection of errors across five commonly used CSS data sets reveals that there is a strong negative relationship between errors of omission and commission, and the probability of making these errors may be associated with the density of the true network. Investigation of the determinants of heterogeneity of perception errors may be beneficial for advancing CSS studies. In the next section we build on divergence of errors in cognition of social relations and introduce a network estimation technique that enables researchers to assess the differential costs of Type 1 and Type 2 errors using a ROC curve based approach.

\section{Network Estimation with CSS}\label{estimation}

The value of being able to estimate the structure of a complete network from a subset of respondents' cognitive social structures is two-fold. First, researchers engaged in a study may not be able to produce a response rate sufficient for network analysis. Unexpected low response rates can result in a failed project or a change in research question. While complete or near complete response rates are typically required for network analysis (Wasserman and Faust, 1994), recent meta-analyses of organizational research found an average response rate of just over 50\%  (Baruch and Holtom, 2008; Anseel et al., 2010).  Second, for researchers interested in cross-organizational research, where the goal is to model both individual and organizational attributes, gathering complete network data from large number of organizations may be impractical and inefficient. Thus, collecting CSS data and using aggregation techniques hold promise for recreating accurate representations of networks when only a sample of data is available.

In this section, given a random sample of CSS slices, our goal is to estimate the whole network by properly aggregating the observed slices. A straightforward approach for this aggregation is to set a threshold, say $k$, add the matrices representing the CSS slices, and if $k$ perceptions have accumulated for a possible tie classify it as an existing tie, otherwise classify as a non-existing tie. Although this seems to be a simple and intuitive approach, there are two important issues. The first issue is the origin of a reported tie, in other words, whether it is a self report or it is a perception on others ties. The second issue is the proper choice of threshold $k$. In this section we first give some details from Siciliano et al. (2012) which considers network estimation by aggregating the randomly selected CSS slices, controlling the Type 1 error rate. We then discuss the main contribution of this paper which is a similar estimation procedure, but controls both Type 1 and Type 2 error rates, therefore considered to be more powerful as well as more flexible.

\subsection{Network Estimation Controlling Type 1 Error}\label{estimation_old}

The first study considering the estimation of a network from randomly sampled individuals' perceptions is Siciliano et al. (2012), where an aggregation method for the observed CSS slices is proposed, so that the Type 1 error rate in the estimated network is kept below a pre-defined tolerable level. This method is shown to have a good estimation power and it forms a basis of our current work, therefore we present the main ideas here. As we have noted above, the first issue regarding the aggregation method is the origin of a reported tie. This is particularly important because as described in Section \ref{types}, the true network is derived through {\em intersection} of all CSS slices, and an estimation procedure is expected to converge to this true network as the sample size increases. Recall that by intersection we mean that a tie between two actors is considered to exist if and only if both actors agree on its existence. Given this, there are three main scenarios that can arise when estimating a certain tie from an observed sample. In the first scenario, both actors of the tie are sampled, therefore existence or nonexistence of this tie will be estimated by the intersection of the self reports from both actors. No perception information will change this. These type of ties will constitute a region in the estimated network matrix referred to as the {\em knowledge region}. In the second scenario, none of the actors of the tie are sampled, therefore, the estimation of this tie only relies on others perceptions. If $k$ perceptions have accumulated in the sample regarding this tie, it is estimated to exist, otherwise, not to exist. In the third scenario, only one of the actors of the tie are sampled, therefore the intersection method cannot be used for estimation. In this scenario, we treat the data from the sampled actor as a perception, and estimate this tie to exist only if $k-1$ additional perception ties are reported by others. Then the estimated ties coming from scenarios 2 and 3 compliment the knowledge region in the estimated network matrix, which is referred to as the {\em perception region}. Now suppose the CSS slices of a random sample of size $n$ are observed from an unknown network. For a given $k$, this aggregation method can be summarized in the following algorithm. In what follows, we will refer to this method as the {\em fixed threshold method} (FTM).

\medskip

{\em FTM Algorithm: CSS Aggregation for a Fixed $k$ }

\begin{enumerate}
\item Estimate the knowledge region of the network using the intersection of self reports (Scenario 1).
\item Estimate the perception region of the network by adding the perceptions and unverified self reports (Scenarios 2 and 3). If there are $k$ reports on a tie, it is estimated to exist.
\item Combine the knowledge and perception parts of the network to get the final estimate.
\end{enumerate}

The second issue regarding the aggregation of CSS slices for network estimation is the selection of a proper threshold $k$. Siciliano et al. (2012) set $k$ such that the estimated Type 1 error rate is controlled below a pre-defined tolerable level $\alpha$. For a given $k$ and an observed sample of CSS slices, one may estimate the Type 1 error rate in the network by

\begin{equation}
\hat{\alpha}_{k}=\frac{
\begin{small}\left\{
\begin{array}{c}
\text{Number of Type 1 errors committed over } \\
\text{the knowledge region with threshold $k$}
\end{array}
\right\}\end{small}
}
{
\begin{small}\left\{
\begin{array}{c}
\text{Number of possible Type 1 errors  } \\
\text{(or the number of zeroes in the knowledge region)}
\end{array}
\right\}\end{small}
}.
\end{equation}

\medskip

Note that this is the same method as we calculate the Type 1 error rate in a complete CSS structure ($n=N$), where the knowledge region is the whole true network observed by intersection method. Since $k$ is inversely proportional to $\hat{\alpha}_{k}$, the smallest $k$ satisfying $\hat{\alpha}_{k}<\alpha$ can be used as a proper threshold value for aggregating CSS slices. Combining this approach with FTM Algorithm, the network estimation method of Siciliano et al. (2012) can be summarized in the following algorithm. As in the original paper, we will refer to this method as the {\em adaptive threshold method} (ATM).

\medskip

{\em ATM Algorithm: Network Estimation from CSS Slices, Controlling Type 1 Error}

\begin{enumerate}
\item Set a tolerable Type 1 error rate $\alpha$. Typical values are 0.05, 0.10, 0.15.
\item Draw a random sample of size $n$ and observe the CSS slices.
\item Find the smallest $k$ such that $\hat{\alpha}_{k}<\alpha$, and denote it by $k^{*}$.
\item Compute the estimated network by aggregating the CSS slices using FTM with threshold $k^{*}$.
\end{enumerate}

\subsection{Network Estimation Controlling Both Type 1 and Type 2 Errors: The ROC Approach}\label{estimation_new}

The estimation methodology described above is somewhat limited in that, it only controls for the Type 1 error rate. As described in Section \ref{types}, another important error type is the Type 2 error, and it is of interest to control both errors when estimating a network from a random sample of CSS slices. Therefore, in this section we propose a new estimation methodology which seeks a balance between Type 1 and Type 2 errors. We base our methodology on a commonly used tool in classification, the receiver operating characteristic (ROC) curve. We first give a brief insight on ROC curves, then we give the specifics of our methodology. We begin by noting that for a given $k$, similar to estimating the Type 1 error rate in the network, it is possible to estimate the Type 2 error rate by

\begin{equation}
\hat{\beta}_{k}=\frac{
\begin{small}\left\{
\begin{array}{c}
\text{Number of Type 2 errors committed over } \\
\text{the knowledge region with threshold $k$}
\end{array}
\right\}\end{small}
}
{
\begin{small}\left\{
\begin{array}{c}
\text{Number of possible Type 2 errors  } \\
\text{(or the number of ones in the knowledge region)}
\end{array}
\right\}\end{small}
}.
\end{equation}

\subsubsection{ROC Curves}

A ROC curve is a graphical plot illustrating the performance of a binary classifier for its varying discrimination threshold parameter. It plots the true positive rate (TPR = {\em Positives correctly classified / Total positives}) against the false positive rate (FPR = {\em Negatives incorrectly classified / Total negatives}) for various levels of the threshold parameter. In classical ROC analysis, typically TPR is plotted on the vertical axis and FPR on the horizontal axis, therefore, the threshold value closest to the top left corner of the plot is said to produce the best classification as it is the value that seeks for larger TPR and smaller FPR. For a detailed treatment of ROC analysis, see, for example, Fawcett (2004).

\subsubsection{ROC Approach for Network Estimation}

Estimating a network with FTM Algorithm is essentially a binary classification problem, where each entry of the network matrix is classified as an existing or non-existing tie, based on the threshold parameter $k$. Here, a Type 1 error instance corresponds to a false positive, and a Type 2 error instance corresponds to a false negative. Similarly, {\em Type 1 error rate} corresponds to FPR, and {\em 1 - Type 2 error rate} corresponds to TPR. Then following the classical ROC approach, the threshold value $k$ closest  to the top left corner of the ROC plot should be the optimal threshold. In other words, when employed along with FTM, the threshold $k$ minimizing the distance

\begin{equation}
\delta=\sqrt{\hat{\alpha}_{k}^{2}+\hat{\beta}_{k}^{2}}
\label{dist}
\end{equation}

gives a network estimate which seeks a balance between Type 1 and Type 2 errors. Note that the threshold value $k$ obtained by minimizing $\delta$ in (\ref{dist}) treats both error types as equally important. However, in practice, one error type may be more crucial than the other, and it is of interest to minimize a weighted distance measure in order to handle these unbalanced cases. In the remaining of this section we first illustrate the ROC based network estimation using the unweighted distance $\delta$ on a real data set and point out potential problems. We then discuss how these problems can be addressed by a ROC approach employing a weighted distance $\delta_{w}$.

\medskip

{\em Numerical Illustration}

\medskip

Consider the High Tech Managers data set introduced in Section \ref{types}, which contains the CSS slices of $N=21$ managers of a machinery firm. Suppose that a random sample of size $n=10$ is drawn and the network is to be estimated from the sampled CSS slices. For this illustration, suppose the individuals 2, 4, 5, 8, 9, 10, 11, 14, 18 and 19 are sampled. We consider FTM for aggregating the slices, and $k$ will be determined by the ROC approach described above, which minimizes the distance $\delta$ in (\ref{dist}). For a given $k$, the true positive rate and the false positive rate in the network may be estimated over the exact region of the network estimate by $1-\hat{\beta}_{k}$ and $\hat{\alpha}_{k}$, respectively. For the observed sample, estimated true positive rates and false positive rates for all possible $k$ are given in Table \ref{tab:illustration}, along with the calculated distance $\delta$. Note that $\delta$ is minimized for $k=1$, therefore FTM with $k=1$ is considered to give a good estimate of the network. However, taking a closer look at the actual numbers of committed Type 1 and Type 2 errors reveal a potential flaw of this approach.

\begin{table}[h]
\begin{center}
\begin{scriptsize}
\centering
\begin{tabular}{ccccc}

$k$	&	TPR ($1-\hat{\beta}_{k}$)	&       FPR ($\hat{\alpha}_{k}$)	&	$\delta$	&	$\delta_{w}$		\\	\hline
0	&	1.000	&	1.000	&	1.000	&	10.606	\\
{\bf 1}	&	0.917	&	0.295	&	{\bf 0.307}	&	3.135	\\
2	&	0.667	&	0.148	&	0.365	&	1.602	\\
3	&	0.583	&	0.080	&	0.424	&	0.941	\\
{\bf 4}	&	0.333	&	0.034	&	0.668	&	{\bf 0.758}	\\
5	&	0.250	&	0.011	&	0.750	&	0.760	\\
6	&	0.083	&	0.011	&	0.917	&	0.925	\\
7	&	0.083	&	0.000	&	0.917	&	0.917	\\
8	&	0.000	&	0.000	&	1.000	&	1.000	\\ \hline

\end{tabular}
\end{scriptsize}
\caption{TPR, FPR, and two distances considered for the numerical illustration.}\label{tab:illustration}
\end{center}
\end{table}

\begin{figure}[h!]
\begin{subfigure}{0.5\textwidth}
  \centering
  \includegraphics[width=1\linewidth]{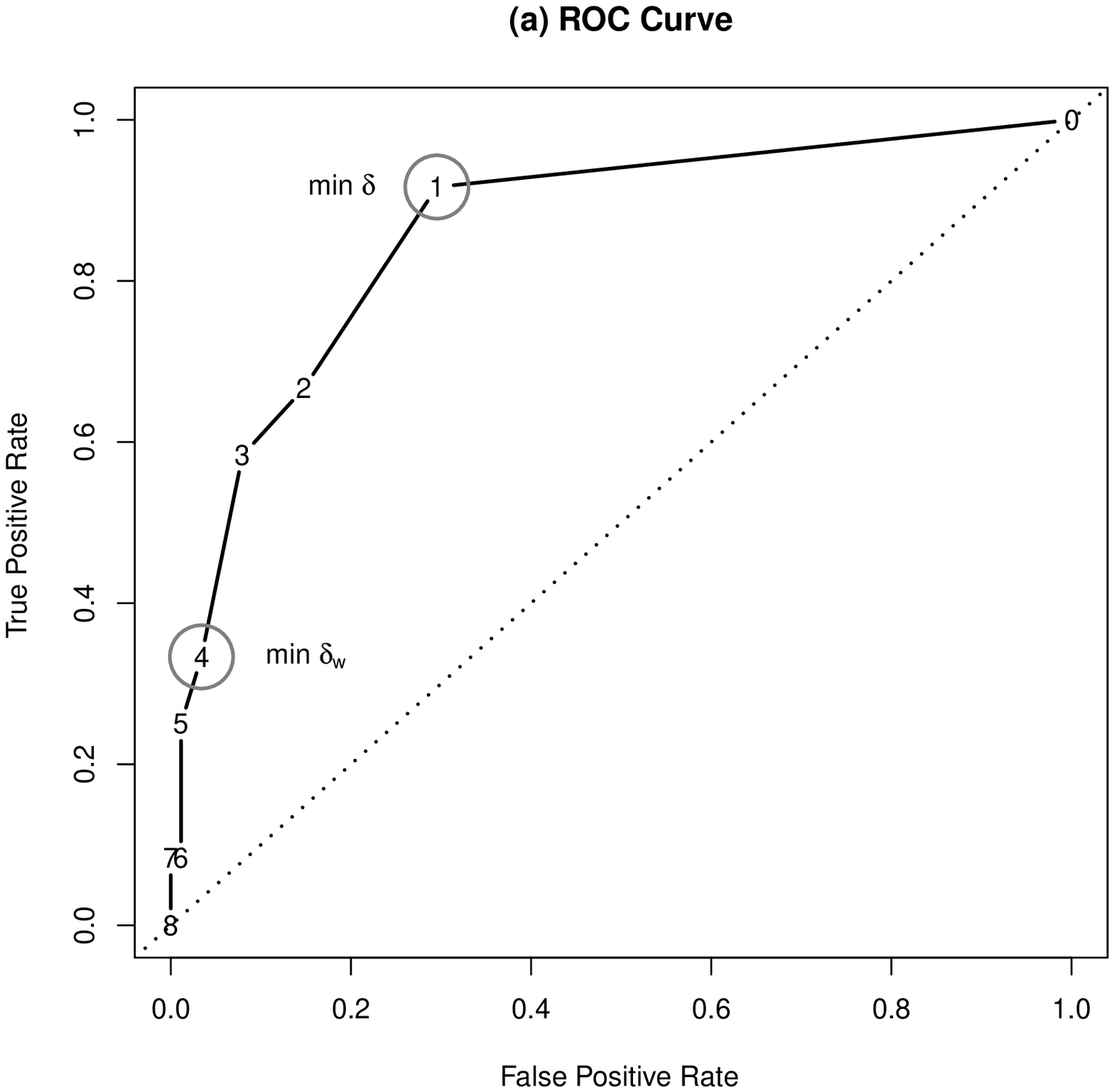}
\end{subfigure}%
\begin{subfigure}{0.5\textwidth}
  \centering
  \includegraphics[width=1\linewidth]{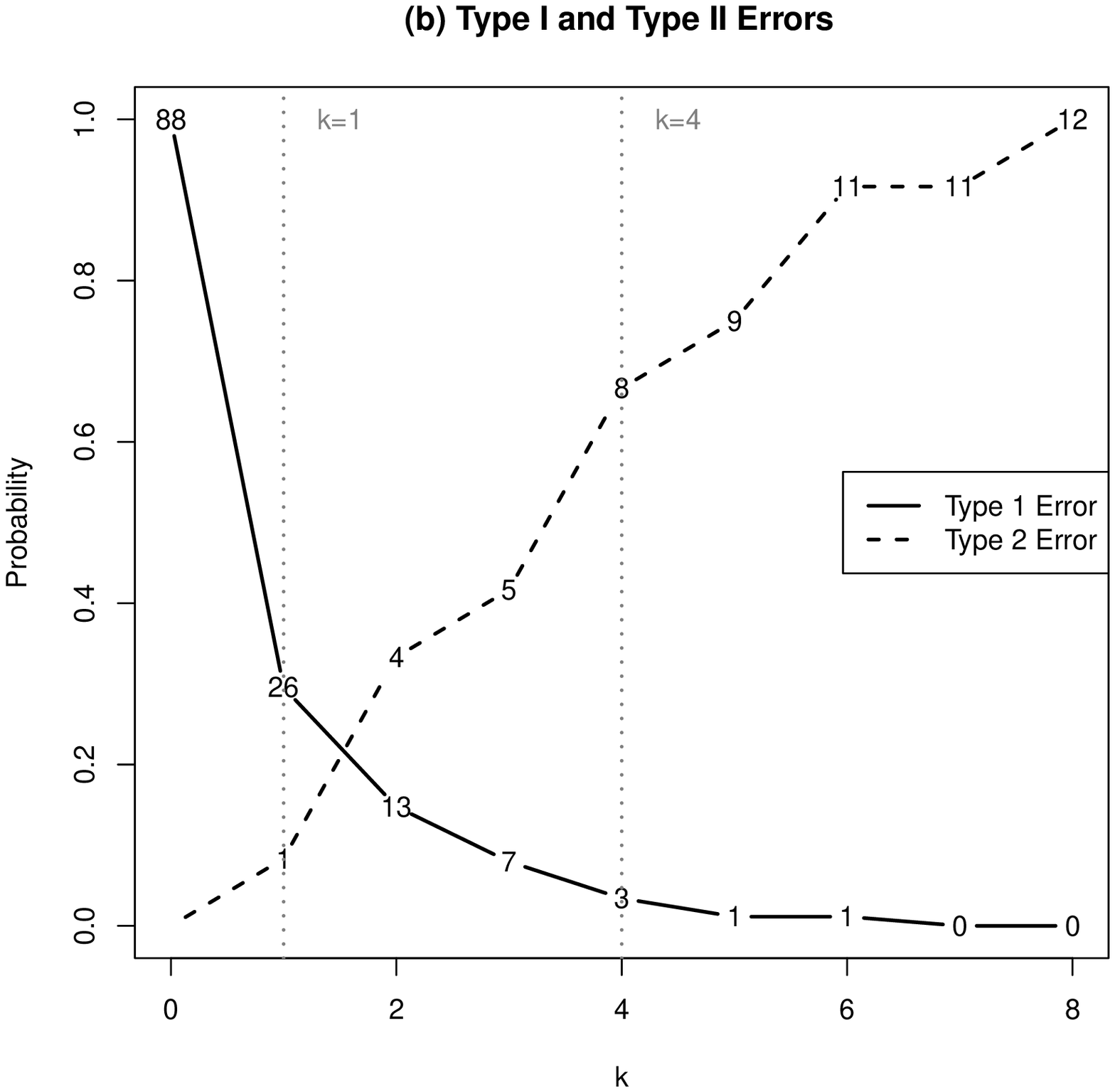}
\end{subfigure}
\caption{Plots for the numerical illustration.}
\label{fig:illustration}
\end{figure}

Figure \ref{fig:illustration}-a illustrates the classical ROC curve produced by varying $k$ from 0 to $8$, along with the value of $k$ at each step. We may see that when $k=1$, the curve is closest to top left corner, i.e., $\delta$ is minimized at this point. Figure \ref{fig:illustration}-b indicates the actual number of Type 1 and Type 2 errors committed for each $k$. We may see that when $k=1$ we commit 26 Type 1 errors out of 88 possible instances ($\hat{\alpha}_{k}=26/88=0.295$), and one Type 2 error out of 12 possible instances ($\hat{\beta}_{k}=1/12=0.083$). The potential drawback here is that the classical ROC approach minimizing  $\delta$ assigns equal importance to both error types, and it gives the optimal $k$ based on error rates, not the actual counts. However, in low density networks reducing Type 1 error may be more critical compared to Type 2 error rate since the true network contains much more zeroes than ones, creating much more potential Type 1 error instances. This can be easily seen in Figure \ref{fig:illustration}-b, where increasing $k$ from one to two reduces the Type 1 error count by 13 at the expense of only three Type 2 error counts, and similarly for the next increment of $k$. Therefore, we consider a weighted distance $\delta_{w}$ to be minimized in the ROC analysis,

\begin{equation}
\delta_{w}=\sqrt{(w\hat{\alpha}_{k})^{2}+\hat{\beta}_{k}^{2}},
\label{distw}
\end{equation}

where a choice of $w>1$ gives more emphasis on Type 1 error, a choice of $w<1$ gives more emphasis on Type 2 error, and the choice of $w=1$ corresponds to classical ROC distance $\delta$. Based on our observation on the effect of network density on the potential error rates, we propose that $w=1/\bar{d}$ is a reasonable choice of $w$ for the low density networks, where $\bar{d}$ is the average density of all CSS slices in the random sample. In the illustrative example $\bar{d}=0.094$, so we set $w=1/0.094=10.606$. As marked in Figure \ref{fig:illustration}-a (as well as in Table \ref{tab:illustration}) , when $w=10.606$ the minimum $\delta_{w}$ is attained for $k=4$, which produces a Type 1 error rate of $\hat{\alpha}_{k}=3/88=0.034$ and Type 2 error rate of $\hat{\beta}_{k}=8/12=0.667$. Then employing FTM with $k=4$ is considered to give a better estimate of the unknown network based on the sample information. As an evidence of the improvement in estimation, we note that when $\delta$ is minimized ($k=1$) the correlation between the estimated network and the true network is 0.644, however, when $\delta_{w}$ is minimized ($k=4$) the correlation is 0.749. Here we also note that the correlation measure we use in this study is the similarity index $S_{14}$ suggested by Krackhardt (1990).

\medskip

{\em Proposed Estimation Method}

\medskip

Based on our observations on the above illustrative example, as well as our investigation on other data sets, we observed that the ROC approach is a powerful one for estimating networks from a random sample of CSS slices since it seeks for a balance between the committed Type 1 and Type 2 errors. We also observed that giving more emphasis on Type 1 error rate improves the estimation in sparse networks. Therefore, we control the level of this emphasis to be inversely proportional to the average density of the observed CSS slices. Our estimation method is summarized in the following algorithm. In what follows, we will refer to this method as the {\em ROC based threshold method} (RTM).

\medskip

{\em RTM Algorithm: Network Estimation form CSS Slices, Controlling both Type 1 and Type 2 Error Rates}

\begin{enumerate}
\item Draw a random sample of size $n$ and observe the CSS slices.
\item Calculate the average density $\bar{d}$ of all CSS slices in the sample, and set $w=1/\bar{d}$.
\item Perform the weighted ROC analysis to find the $k$ such that $\delta_{w}$ is minimized, and denote it by $k^{*}$.
\item Compute the estimated network by aggregating the CSS slices using FTM with threshold $k^{*}$.
\end{enumerate}

As will be illustrated in the following section, our numerical study across five data sets indicate that setting the weight to $w=1/\bar{d}$ is a reasonable choice. However, for various reasons, a researcher might want to use a different weight, and our ROC based methodology may be employed for any choice of $w$. This may be needed for certain social networks such as terrorist networks in which costs of Type 2 errors may be considered to be significantly larger than Type 1 errors. We finalize this section by noting that FTM, ATM, and RTM algorithms may be implemented by the functions  \texttt{ftm},  \texttt{atm} and  \texttt{rtm}, respectively, in the \texttt{cssTools} package for the statistical software R. The illustrative High Tech Managers data set is also included in the package, along with functions for producing tables and figures similar to Table \ref{tab:illustration} and Figure \ref{fig:illustration}, and some useful tools for the analysis of CSS data.

\section{Numerical Study}\label{numerical}

\begin{figure}[h]
\begin{center}
\includegraphics[scale=0.62]{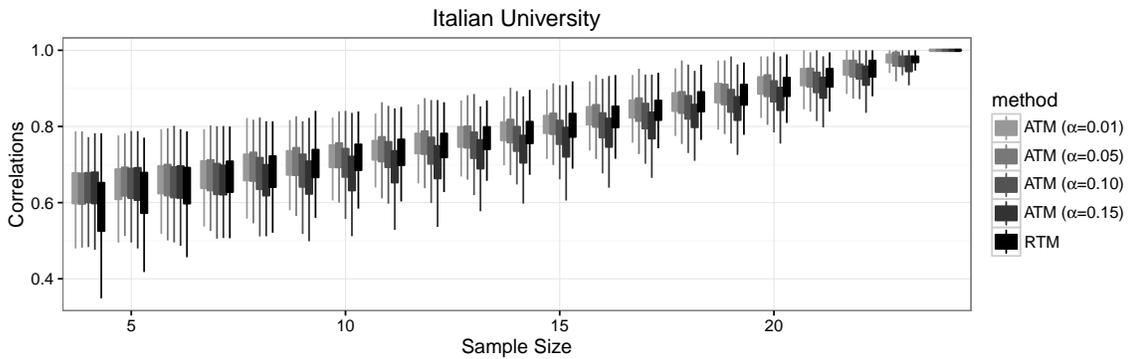}
\caption{Italian University data. Correlations between true and estimated networks.}
\label{fig:italian_cor}
\end{center}
\end{figure}

\begin{figure}[h]
\begin{center}
\includegraphics[scale=0.62]{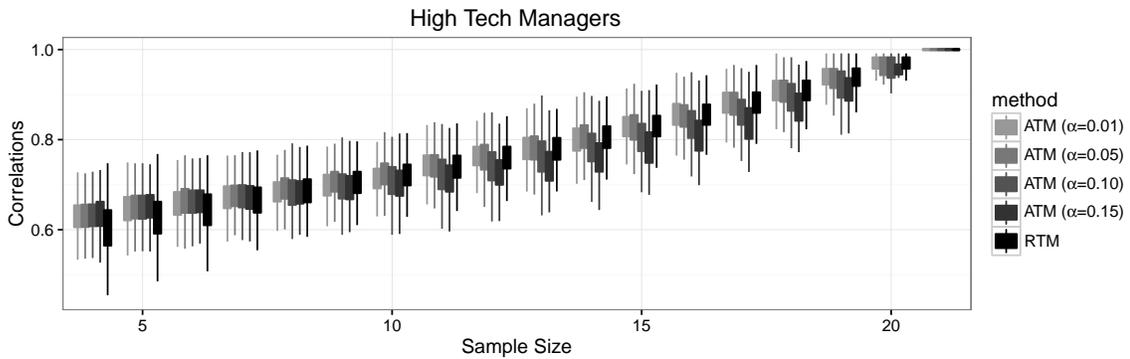}
\caption{High Tech Managers data. Correlations between true and estimated networks.}
\label{fig:hightech_cor}
\end{center}
\end{figure}

\begin{figure}[h]
\begin{center}
\includegraphics[scale=0.62]{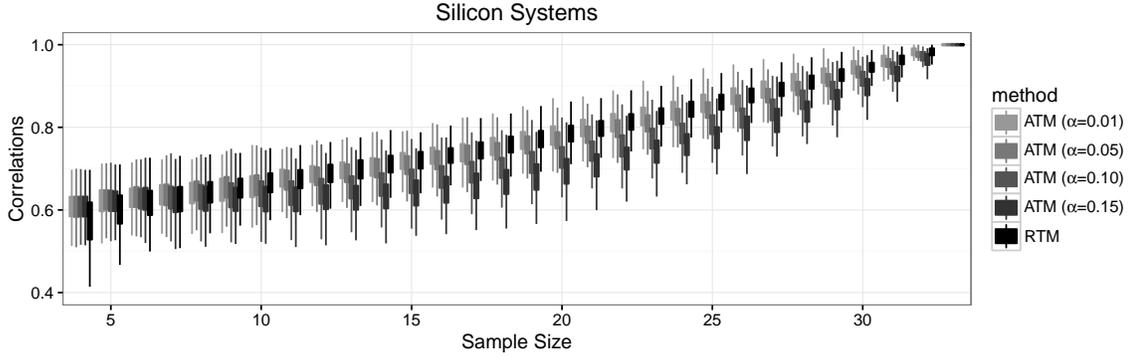}
\caption{Silicon Systems data. Correlations between true and estimated networks.}
\label{fig:silicon_cor}
\end{center}
\end{figure}

\begin{figure}[h]
\begin{center}
\includegraphics[scale=0.62]{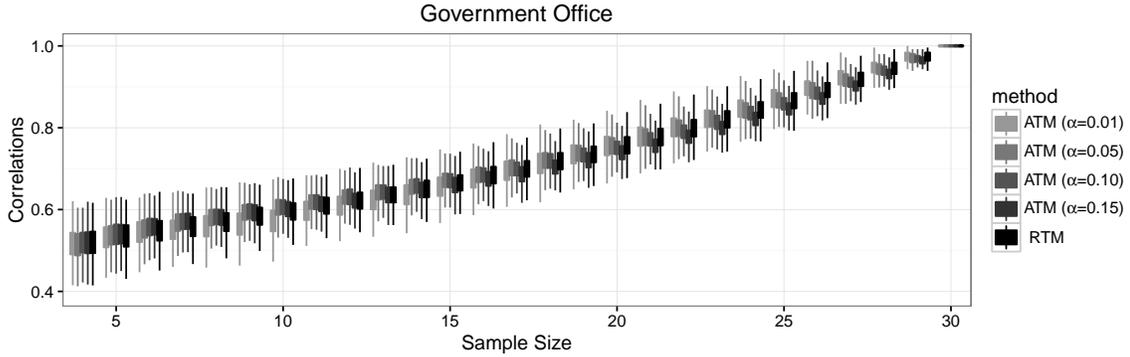}
\caption{Government Office data. Correlations between true and estimated networks.}
\label{fig:government_cor}
\end{center}
\end{figure}

\begin{figure}[h]
\begin{center}
\includegraphics[scale=0.62]{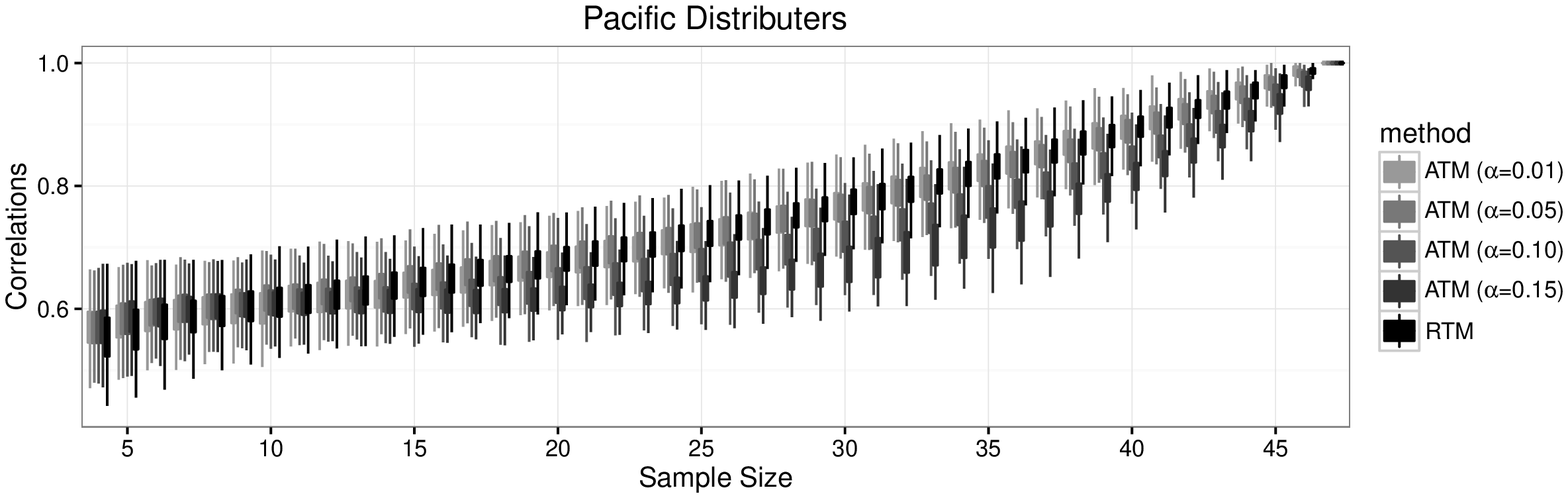}
\caption{Pacific Distributers data. Correlations between true and estimated networks.}
\label{fig:pacific_cor}
\end{center}
\end{figure}

In this section we present the results of a numerical study aiming to illustrate the performance of the proposed estimation methodology. Since the adaptive threshold method (ATM) of Siciliano et al. (2012) for estimating networks from a random sample of CSS slices is known to outperform traditional roster and ego network methods, we only compare the proposed ROC based adaptive threshold methodology (RTM) with ATM. For each data set introduced in Section \ref{types}, we generate random samples of sizes 4 to $N$, estimate the network based on the sample information only, and compute the correlation ($S_{14}$) between the estimated network and the true network obtained by the intersection method. Our results across five data sets are displayed in Figures \ref{fig:italian_cor} to \ref{fig:pacific_cor}, which display the boxplots of the observed correlations for different estimation methods. In each plot, the horizontal axis represents the sample size, the vertical axis represents the correlation, and the boxplots illustrate the distribution  of the observed correlations based on 1000 simulations. In other words, for each sample size the vertical boxes represent the middle 50$\%$, and the vertical lines represent the lower and upper 25$\%$ of the distribution of the observed correlations between the true and estimated networks. The colors indicate the method used for estimation.

We may see from the figures that the overall performances of both methods considered seem to be satisfactory. As expected, for all data sets and methods, the correlation between the estimated and true networks increase as the sample size increases, and when all individuals are included in the sample the correlation equals one. The performances of the two methods considered seem to be comparable, however, ATM turns out to be sensitive to the selection of tolerable Type 1 error rate $\alpha$. When using ATM, setting $\alpha$ to a larger value such as 0.15 generally seem to be a poor selection, which is most visible for Silicon Systems and Pacific Distributers data sets (Figures \ref{fig:silicon_cor} and \ref{fig:pacific_cor}, respectively). This is an expected result as too much Type 1 error is allowed in the estimation procedure. On the other extreme, setting $\alpha$ to a smaller value such as 0.01 seems to produce better results across most data sets, however, as we see in Pacific Distributers data for small to moderate sample sizes, choice of 0.15 may outperform choice of 0.01. This may be because of the fact that forcing to reduce Type 1 error to a great extent may trigger more Type 2 errors and thus result in a larger overall error rate. Across the five data sets considered, setting $\alpha$ to 0.05 or 0.10 seem to be safer, but this feature cannot be generalized to any data set. In short, a researcher to employ ATM may be advised to set $\alpha$ to 0.05 or 0.10, but, they must also be advised that these are the preferable levels for the five data sets investigated, and it is not guaranteed that they will be the best choices for any data set.

On the other hand, the proposed RTM performs as good as ATM with a good choice of $\alpha$, and it provides consistent results across all data sets. In this sense it is a robust method for estimating networks, and the fact that it doesn't require a pre-determined decision threshold such as $\alpha$ is appealing. RTM has an automatic way of balancing Type 1 and Type 2 errors, and our numerical study shows that its estimation performance is comparable with ATM, even when the ATM user is assumed to choose the best $\alpha$ level.

\section{Conclusion}\label{conclusion}

We first presented the common patterns of errors across five CSS data sets. Consistent in all data sets is the strong negative association between Type 1 and Type 2 errors. For the individuals with lower overall errors, most of the errors are of Type 2, whereas for the individuals with higher overall errors most of the errors are of Type 1. Due to low density of the true networks considered, a common feature in many social networks, the uninformed perceivers tend to make much more Type 1 errors. These findings have especially important implications for measurement of network data, hence investigation of the determinants of tendency for different error types may be a useful next step in CSS research.

Next, we presented a novel estimation method, RTM, which requires a sample of CSS slices from the network of interest, and returns an estimate of the network while controlling for both Type 1 and Type 2 errors. RTM utilizes the ROC curve for balancing two error types, and it does not require a user decision on any cut-off values (such as $\alpha$ of ATM). Our comparative numerical study shows that the proposed method performs well. RTM gives weights to error types based on the network density, however, researchers may also choose to set their preferred weights depending on the emphasis they need to give to error types depending on their consequences.

Using RTM, it is possible to incorporate network data collection to random survey studies covering multiple organizations. Likewise more scholars are recognizing the need to study organization networks at multiple levels (Brass et al. 2004; Zappa and Lomi, 2015;  Tranmer et al., 2016), however, data collection requirements are extremely demanding for studying inter-organizational and intra-organization networks simultaneously. RTM provides a convenient approach to meet data requirements for multi level organization research as well. One limitation of RTM is the network size. Due to obvious individual cognitive limitations, it may not be wise to apply RTM to networks involving  more than 70 individuals. For large networks, a clustered version of RTM that is based on major departments or units in organizations can be considered.

Despite the promise of random sampling, gathering complete network data on the actors under study remains the best and recommended approach for collecting network data. We see these methods as useful when a researcher expects or obtains a low response rate or when the scope of the study prohibits the collection of network data from all participants, such as one looking at networks in a large number of organizations.


\end{document}